# Davydov Splitting, Resonance Effect and Phonon Dynamics in CVD grown Layered $MoS_2$

Deepu Kumar[1#], Birender Singh[1], Rahul Kumar[2], Mahesh Kumar[2], Pradeep Kumar[1*]

[1]*School of Basic Sciences, Indian Institute of Technology Mandi, 175005, India*

[2]*Department of Electrical Engineering, Indian Institute of Technology Jodhpur, 342037, India*

We present a comprehensive temperature dependent Raman measurements for horizontally aligned CVD grown layered $MoS_2$ in a temperature range of 4 to 330 K under resonance condition. Our analysis of temperature dependent phonon frequency shift and linewidth suggest a finite role of three and four phonon anharmonic effect. We observed Davydov splitting of the out-of-plane ( $A_{1g}$) and in-plane ($E_{2g}^1$) modes, attributed to the weak interlayer interaction, and reflected in the appearance of additional modes with decreasing temperature for both 3 layers and few layers system. We also observed that the number of Davydov splitting components are more in few layers as compared to 3L $MoS_2$, suggesting it increases with increasing number of layers. Temperature evaluation of the Raman spectra shows that the Davydov splitting, especially for $A_{1g}$ mode, is very strong and well resolved at low temperature. We note that $A_{1g}$ mode shows splitting at low temperature, while $E_{2g}^1$ mode is splitted even at room temperature, and that suggests to prominent role of $A_{1g}$ mode to the interlayer interaction. Further, the temperature dependence tuning of resonance effect is observed, via almost sixty fold increase in the intensity of the phonon modes at low temperature.

[#]E-mail: deepu7727@gmail.com

*E-mail: pkumar@iitmandi.ac.in

## 1. Introduction

Two dimensional layered transition metal di chalcogenide (TMDCs) materials have been focused of research community due to their unique properties and potential applications in electronic and optoelectronic devices. The TMDCs have a general formula $MX_2$, where M and X correspond to transition metal and chalcogen atom, respectively; and a single layer is composed of a sheet of transition metal atom sandwiched between two sheets of chalcogen atoms in a form of X-M-X either with trigonal prismatic hexagonal or octahedral structure [1-8]. In these materials, atoms within the layer are strongly bounded with strong covalent bond, while the interaction between the layers is weak van der Waals type. The electronic and vibrational properties of single layer are significantly different from the bulk, for example bulk $MX_2$ has indirect band gap, which changes from indirect to direct as changing of thickness from bulk to monolayer and as results a significant enhancement in the photoluminescence [9-11], large carrier mobility [12-13] make these an attractive materials for flexible field effect transistor as well as for other electronic and optoelectronic devices.

In the present work, we report a comprehensive study of temperature dependent vibrational properties for horizontally aligned CVD grown layered $MoS_2$ by using Raman spectroscopic technique under resonance condition. Raman spectroscopy is a non-destructive and fast technique and has been widely used to explore the vibrational and electronic properties of a wide range of materials [14-19]. $MoS_2$ is one of the early member of TMDCs family and is very crucial from fundamental as well as application point of view. Photoluminescence spectra for very thin layer of $MoS_2$ shows two features in the excitonic energy range from ~1.8 to ~2.1 eV [20-22]. Splendiani et al. shows photoluminescence spectra for monolayer $MoS_2$ consisting of two peaks centered at 1.83 eV and 1.98 eV and absorption spectra shows two peaks centered at 1.85 eV and 1.96 eV attributed to A and B excitons [22]. When excitation laser energy is very close to the allowed excitonic energy state, the resonance effect occurs and influences the Raman scattering intensity of phonons and it results into several strong peaks in the Raman spectra of $MoS_2$, which are not observed under non-resonance condition. In non-resonance, only the first-order Raman scattering is dominated i.e. Brillouin zone center phonons participate in the scattering process, while under resonance condition, phonons with non-zero wave vector also participate in the Raman scattering and gives rise to strong second-order Raman scattering. It has been previously reported that Raman spectra of $MoS_2$ and other 2D materials show strong dependence on the laser excitation sources [23-26]. In addition to the enhancement of the Raman spectra, splitting in some of the first-order phonon modes has also been observed in few layered system such as $WS_2$, $MoSe_2$, $MoTe_2$ under resonance condition, and is attributed to the Davydov splitting [27-30]. We note that in case of $MoS_2$, only one report, and that too only at room temperature has mentioned about the splitting of modes as a function of layer number attributed to Davydov splitting [26]. Davydov splitting in optical phonon modes appears due to weak van der Waals interaction between the layers [28]; and as this splitting is directly related to interlayer interaction, hence expected to be dependent on the temperature.

Temperature dependent Davydov splitting can affects the strength of interlayer interaction which may be reflected via the phonons. Most of the physical properties such as optical, vibrational and electronic, of these 2D materials are layer dependent and hence intriguingly related to the interlayer interaction. In addition, self-heating of device based on these 2D materials and electron-phonon interactions control the mobility of charge carriers resulting in reduced performance of the devises. As mobility of the charge carriers are associated with the electron-phonon interaction, therefore any changes in the strength of electron-phonon interaction will affect the performance of electronic devices. Hence, it become important to investigate the strength of interlayer interaction and detailed phonon dynamics of $MoS_2$ as well as other similar 2D materials for understanding dependence of physical properties on the number of layers as well as on the temperature.

Here, we have undertaken a detailed temperature dependent studies on horizontally aligned CVD grown layered $MoS_2$ in a temperature range of 4 to 330 K. In the present work, we have used 633 nm (~1.96 eV) laser as an excitation source, which is close to the B exciton energy state in $MoS_2$, and therefore it resonates with excitonic energy state, and as a result a large number of the phonon modes are observed along with first order Raman active modes. Our comprehensive temperature dependent studies revel anomalies in the phonon modes reflected via the renormalization of the phonon self-energy parameter (i.e. mode frequencies and linewidths) as well as appearance of additional modes with lowering temperature attributed to the Davydov splitting suggesting the tunability of interlayer interaction with temperature.

## 2. Experimental Details

Horizontally aligned $MoS_2$ samples were synthesized using CVD method as described in Ref. [31]. Temperature dependent Raman measurements were carried out using Horiba LabRAM HR evolution Raman Spectrometer in the backscattering geometry. The sample was illuminated with 633 nm line of laser using a 50x long working distance objective lens to both focus laser beam on the sample and for collecting the scattered light from the sample. The laser power on the sample was kept very low ≤ 1 mW to avoid any heating effect. The scattered light was detected by using 1800 grove per mm grating coupled with Peltier cooled Charge Coupled Device detector. The temperature variation was carried out using closed cycle refrigerator (Montana) in a temperature range from 4 to 330 K with temperature accuracy of ± 0.1 K.

## 3. Results and Discussions

### 3.1. Phonon Raman scattering for 3L and few layers $MoS_2$

Bulk $MoS_2$ belongs to the space group $P6_3/mmc,\#194$ and the point group $D_{6h}^4$, and within the unit cell there are 6 atom per unit cell giving 18 normal modes of vibrations at the $\Gamma$ point of the Brillouin zone (BZ) with the irreducible representation given as $A_{1g}+2A_{2u}+2B_{2g}+B_{1u}+E_{1g}+2E_{1u}+2E_{2g}+E_{2u}$ [32-34]. There are four first-order Raman active modes in bulk $MoS_2$, namely $A_{1g}$, $E_{1g}$, $E_{2g}^1$ and $E_{2g}^2$ .The first-order Raman active phonon mode $A_{1g}$ appears due to the out-of-plane vibrations of S atoms in opposite directions, and the mode $E_{2g}^1$ corresponds to the in-plane vibrations of Mo and S atoms in opposite directions. Raman active mode with $E_{1g}$ symmetry is an in-plane mode which arises due to in-plane vibrations of S atoms in opposite directions and generally it is absent in the backscattering geometry. Raman active mode with $E_{2g}^2$ symmetry appears due to vibration of the adjoining layers and therefore absent in monolayer. Monolayer/odd number layers of $MoS_2$ belongs to point the group $D_{3h}^1$ and the space group $P\bar{6}m2,\#187$, there are 3 atoms per unit cell giving 9 normal modes of vibrations at the $\Gamma$ point of the BZ with irreducible representation as $A_1^{'}+2E^{'}+2A_2^{''}+E^{''}$ [32-34]. However, bilayer/even number layers of $MoS_2$ belongs to the point group $D_{3d}^3$ and the space group $P\bar{3}m1,\#164$, there are 6 atoms per unit cell giving 18 normal modes of vibrations and can be expressed into the following irreducible representation $3A_{1g}+3E_g+3A_{2u}+3E_u$ [32-34]. We note that in case of even/odd layers, phonon modes notation changes. For simplicity, we have followed notation of bulk inline with literature [35-36]. Raman spectroscopic technique can be used as a fingerprint to identify few number of layers (1-6) present in TMDCs via qualitatively determining the frequency difference between the $A_{1g}$ and $E_{2g}^{1}$ phonon modes [37]. For CVD grown monolayer $MoS_2$, the frequency difference between $A_{1g}$ and $E_{2g}^{1}$ phonon mode is ~20 $cm^{-1}$ [38-39], which is ~ 2 $cm^{-1}$ larger than that for mechanically exfoliated $MoS_2$ [34-35]. There is an increment of ~2-3 $cm^{-1}$ when one goes from monolayer to bilayer, and after that, difference increase by ~1 $cm^{-1}$ with addition of one layer till 5-6 layers, however for bulk system, this frequency difference is ~ 25 $cm^{-1}$ [34-35, 37-39].

Figure 1 (a) shows the optical micrograph for horizontally aligned CVD grown layered $MoS_2$. The region where temperature dependent Raman measurements were carried out under resonance condition are labeled as (1) and (2) corresponding to the flake 1 and flake 2, respectively. To identify the number of layers, we first excited spectra under non-resonant condition (i.e. using 532 nm (2.33 eV) laser) at flake 1 and flake 2 (see Fig. 1 (a)), and the collected Raman spectra are shown in Fig. 1 (b). The first-order Raman active phonon modes $E_{2g}^{1}$ and $A_{1g}$ are observed at 382.2 $cm^{-1}$, 381.6 $cm^{-1}$; 405.8 $cm^{-1}$, 407.0 $cm^{-1}$ in flake 1 and flake 2, respectively. The obtained value of frequency difference between $A_{1g}$ and $E_{2g}^{1}$ is ~ 23.6 $cm^{-1}$ for flake 1, suggesting 2-3 layers, while it is ~ 25.3 $cm^{-1}$ for flake 2, indicating multiple number of layers. To confirm number of layers, we also performed photoluminescence (PL) measurements. Figure 1 (d) shows the PL spectra collected from the flake 1 and flake 2. Flake 1 shows two strong PL features centered at 1.83 eV and 1.97 eV corresponds to A and B exciton, and these values are very close to the absorption peak as reported for 3L $MoS_2$ [40], which again confirmed that flake 1 consists not more than 3 layers. Flake 2 shows a very weak PL signal suggesting that flake 2 is not a bulk but approximately have 6-7 layers, and for convenience, we have adopted nomenclature few layers $MoS_2$ in the whole manuscript.

Figure 2 shows the Raman spectra for $MoS_2$ at 4 K in spectral range of 360-700 $cm^{-1}$, under resonant excitation i.e. using 633 nm (1.96 eV). The bottom and top panel in Fig. 2 show the Raman spectra for few layers and 3L $MoS_2$, respectively. Both the spectra, few layers as well as 3L $MoS_2$ show the first-order as well as a large number of second-order vibrational modes. The spectra are fitted with the sum of Lorentzian functions to extract spectral parameters such phonon mode frequency ($\omega$), full width at half maximum (FWHM) and the integrated intensity of the individual phonon modes. We have labeled all the observed modes as S1-S20 for convenience. Table-I illustrates the frequencies of the observed modes at 4 K as well as 300 K along with corresponding symmetry assignment of the modes for few layers as well as 3L $MoS_2$. The mode assignment is done in accordance to the earlier reports on $MoS_2$ [41-43]. Figure 2 (b) shows the Raman spectra for few layers in the spectral range of 360 -700 $cm^{-1}$ at 4 K. Under resonance condition, the first-order phonon mode S2 ($E^1_{2g}$) and S6 ($A_{1g}$) appeared at 384.9 $cm^{-1}$ and 410.0 $cm^{-1}$ in the spectrum. Also, apart from these first-order phonon modes, we noticed a few additional modes attributed to the Davydov splitting, discussed in later section. We observed a prominent peak S7 at ~ 421.5 $cm^{-1}$, which is very close to second order dispersive mode as observed for bulk $MoS_2$ [43] under resonance condition, and it is related to resonant second-order Raman process and can be assigned as combination of the modes involving a dispersive quasi-acoustic phonon ($B^2_{2g}$) and a dispersionless phonon ($E^2_{1u}$). We note that Sekine et al. [44] interpreted this highly dispersive mode to a two-phonon Raman scattering process and understood it as follow: (i) First a polariton is created by an incident photon in the inner branch. (ii) Then the polariton is scattered to the outer branch by emitting a longitudinal quasi-acoustic phonon with large wave vector parallel to the c axis. (iii) And finally, polariton is scattered by emitting an optical phonon into the photon like final state. Furthermore, interestingly we notice the emergence of a new mode S8, at low temperature around ~ 150-160 K, with decreasing temperature, and on further cooling this mode becomes prominent and is clearly separated (see Fig. 3 (a)). The origin of this mode is still not understood. In the spectral range of 435-490 cm$^{-1}$, a broad and intense asymmetric feature is observed at frequency ~ 460 cm$^{-1}$, previously it was assigned to the second-order longitudinal acoustic ($2LA(M)$) phonon mode [41], and later Frey et al. [42] suggested that this broad asymmetric band shows double mode feature, and attributed to the combination of $2LA(M)$ and first order forbidden IR active phonon mode ($A_{1u}$) from the $M$ and $\Gamma$ point of the Brillouin zone, respectively. Here, we have fitted this broad asymmetric peak with four Lorentzian peaks for best fitting to the experimental data. We have assigned peak S10 centered at ~ 456.8 cm$^{-1}$ to the longitudinal acoustic mode ($2LA(M)$) and a shoulder mode at ~ 470.5 cm$^{-1}$ (S12) to the IR active phonon mode ($A_{1u}$). However, recent reports attributed different symmetry assignment for modes contributing to this broad asymmetric peak [45-46]. Livneh et al. [45] considered an additional peak centered at ~ 440 cm$^{-1}$ and suggested that this peak arises due to second order Raman scattering and is combination of interlayer-rigid ($E_{2g}^2$) and out-of-plane ($A_{1g}$) modes. Furthermore, the spectral range of 510-700 cm$^{-1}$, is dominated by second-order phonon modes which appear due to multi-phonon Raman scattering process and it can be assigned to the combination of modes involving optical phonons at the $M$ point coupled to the $LA(M)$ mode. The second-order modes are observed at 528.7 cm$^{-1}$ ($E_{1g}(M)+LA(M)$; S13), 573.3 cm$^{-1}$ ($2E_{1g}(\Gamma)$; S16), 602.9 cm$^{-1}$ ($E_{2g}^1(M)+LA(M)$; S18) and 645.8 cm$^{-1}$ ($A_{1g}(M)+LA(M)$; S20). Moreover few additional modes are also observed at 545.4 cm$^{-1}$ (S14), 562.0 cm$^{-1}$ (S15), 591.4 cm$^{-1}$ (S17) and 635.8 cm$^{-1}$ (S19). The origin and symmetry assignment of these modes is not yet clear but one may expect that these modes also appeared due to multi-phonon Raman scattering which are more prominent under resonance condition. Figure 2 (a) shows the Raman spectra for 3L $MoS_2$ in the spectral range of 360-700 cm$^{-1}$ at 4 K. We have labeled all the observed mode as S1- S20 for convenience. We note that this numbering of modes for 3L $MoS_2$ has been done in line with to the few layers $MoS_2$ (see Fig. 2 (b)). First order Raman active mode $E^1_{2g}$ (S2) and $A_{1g}$ (S6) are observed at frequency at 385.6 $cm^{-1}$ and 408.8 $cm^{-1}$, respectively. Similar to the few layers, second-order dispersive mode S7 is observed at ~ 424.5 $cm^{-1}$. The observed second-order modes and their symmetries assignment are listed in Table-I.

## 3.2. Temperature dependent splitting of the phonon modes

Besides the observations of a large number of phonon modes corresponding to the first- as well as second-order Raman scattering under resonance condition, we also observed splitting of the first-order optical phonon modes attributed to weak interlayer interactions i.e. Davydov splitting. Davydov splitting is generally observed in A-like vibration modes i.e. out-of-plane modes [28-29], which intrinsically coupled with the interlayer interaction. In N layer $MX_2$ $(N \geq 2)$, the $A_{1g}$ mode is splitted into N Davydov components. These splitted components are either Raman or Infrared active modes. In N layer $MX_2$ $(N \geq 2)$, the $A_{1g}$ mode splits into $\frac{N}{2} A_{1g}$, $\frac{N+1}{2} A_{1g}$ Raman active modes and $\frac{N}{2} A_{2u}$, $\frac{N-1}{2} A_{2u}$ Infrared active modes for even and odd number of layers, respectively [29]. The total number of Dvaydov splitting is equal to the number of layers, therefore Davydov splitting of $A_{1g}$ mode is another way to estimate number of layers in $MX_2$ system. Davydov splitting is more prominent and visible for a specific wavelength which is closer to the excitonic energy state and hence occurring of such kind of splitting also reflects the resonance effect. Since Davydov splitting is directly related to the interlayer interaction, therefore the strength of the interlayer interaction can be understood by analyzing the splitting. Figures 3 (a) and 3 (b) show the temperature evaluation of the Raman spectra, and clearly showing the mode splitting in the spectral range of 360-435 $cm^{-1}$ for few layers and 3L $MoS_2$, respectively. At higher temperature (i.e. 320 K), $A_{1g}$ (S6) mode in few layers shows an additional weak component as a shoulder mode. With decreasing temperature, splitting becomes more prominent and resolvable, and in addition to shoulder mode, a new mode also start emerging near ~160 K, and these become stronger and well resolved with further decrease in temperature (see Fig. 3 (a)). For the case of 3L $MoS_2$, at the higher temperature, we observed only $A_{1g}$ (S6) mode in the spectral range of 400-412 $cm^{-1}$. With decreasing temperature, $A_{1g}$ mode is splitted into two components around 270 K and these become distinct with further decrease in temperature, see mode S4 at 4 K in Fig. 3 (b). We note that such temperature dependent Davydov splitting of $A_{1g}$ mode, shown here for few layers and 3L $MoS_2$, was also reported recently for the case of $WS_2$ [25]. Now we will turn to the splitting of in-plane mode ($E^1_{2g}$). For few layers $MoS_2$, we observed one broad mode (S1) and one sharp mode (S3) near $E^1_{2g}$ (S2). These modes have been attributed to the Davydov splitting of the mode $E^1_{2g}$ [43, 45]. These three modes are well resolved at higher temperature and as temperature decreases, these modes began to overlapping in contrast to single 2H-$MoS_2$ bulk crystal. [47]. Our results show that mode S2 is ~ 3 times stronger than mode S3 at the room temperature, and with lowering temperature, S3 starts to gain spectral weight and become prominent at 4 K. Similar to few layers $MoS_2$, splitting of in-plane mode $E^1_{2g}$, is also seen in 3L $MoS_2$ (see Fig. 3 (b)). We observed an additional mode S1 in the lower frequency side of $E^1_{2g}$ (S2) mode. We note that Davydov splitted modes, especially for $A_{1g}$ show strong intensity enhancement with decreasing temperature, see Figs. 3 (a) and 3 (b). This effect is clearly seen in the 2D color contour map of the intensity verses Raman shift against temperature as shown in Figs. 3 (c) and 3 (d) for few layers and 3L $MoS_2$, respectively. Such an anomalous increase in the intensity with decreasing temperature may be understood as: the interlayer spacing between the layers decreases with decrease in temperature resulting into increases in the strength of interlayer coupling, and is reflected as enhanced Davydov splitting. Also, the other possible reason to the enhancement in Raman scattering intensity of the modes at low temperature may arise due to the tuning of resonance effect from B exciton near room temperature to A exciton at low temperature, and will be discussed in later section. We also observed that Davydov splitting for both few layers as well as 3L $MoS_2$ (see Fig. 1 (b)), is not observed when the Raman spectra is excited with 532 nm (2.33 eV) laser suggesting that such kind of splitting is usually very weak and is observed only under resonance condition.

## 3.3. Temperature dependence of the phonon modes

To understand the temperature evaluation of the phonon self-energy parameters such as frequency ($\omega$) and full width at half maximum (FWHM), we extracted frequency ($\omega$) and FWHM of the phonon modes via fitting with the multiple Lorentzian function and plotted as a function of temperature. Figures 4 (a) and 4 (b) illustrate the mode frequency ($\omega$) and linewidth (FWHM) of the few prominent modes as a function of temperature for few layers $MoS_2$ in the spectral range of 360-510 $cm^{-1}$. Following observations can be made from our comprehensive temperature dependent Raman measurements: (i) Frequency of all the observed modes S2 ($E_{2g}^1$), S4-S6 ($A_{1g}$), S7, S8, S12 and S1/S3/S10/S11 (not shown here) show normal behavior i.e. mode softening with increasing temperature. (ii) Linewidth of the modes S6 ($A_{1g}$), S7 and S1/S4 (not shown here) show normal temperature dependence i.e. linewidth increases with increasing temperature. The changes in the linewidth of the mode S2 ($E_{2g}^1$) as a function of temperature is very small. Temperature dependence of linewidth of modes S3/S10 (not shown here) show similar behavior as seen for S2 ($E_{2g}^1$). Linewidth S11 (not shown here) and S12 show normal behavior from 4 to ~ 240 K; However, above ~ 240 K it shows anomalous temperature dependence i.e. start decreasing with increasing temperature.

Figures 4 (c) and 4 (d) illustrate the mode frequency ( $\omega$ ) and linewidth as a function of temperature for few layers $MoS_2$ in the spectral range of 510-700 $cm^{-1}$. Following observations can be made: (i) Frequency of the modes S16, 14/S17 (not shown here) and S18-S20 show normal temperature dependence. The mode S15 (not shown here) show anomalous behavior from 4 to ~ 80 K; however, above ~ 80 K it show normal behavior with increasing temperature (ii). Linewidth of mode S14 (not shown here), S16 and S20 show normal temperature dependence above ~160 K and below this, linewidth of the mode S16 remain nearly temperature independent, while linewidth for the modes S14/S20 slightly increases with decreasing temperature. Linewidth of mode S19 shows normal behavior in the temperature range of 4 to ~ 80 K, and above 80 K it shows anomalous behavior i.e. first linewidth decreases with increase in temperature from ~ 80 to 160 K and above this it remains nearly constant till highest recorded temperature. The behavior of linewidth of mode S18 is rather peculiar, it is nearly constant in the temperature range of 4 to 40 K and above 40 K it shows normal behavior in the temperature range of ~ 40 to 100 K. With further increase in temperature, it decreases till ~ 160 K and then start increasing till ~ 200 K and above this temperature, it again start decreasing with increasing temperature. Linewidth of mode S17 (not shown here) shows normal behavior in the temperature range of 4 to 100 K then it decreases with increasing temperature till ~160 K and with further increase in temperature it shows normal behavior. Similar anomalies in the linewidth are also reported for different families of TMDCs [48-49] attributed to the mismatch between expansions coefficients of the underlying substrate.

Figures 5 (a-d) illustrate the mode frequency and linewidth of the few prominent modes as a function of temperature for 3L $MoS_2$. Following observation can be made: (i) Frequencies of all the observed modes (few modes are not shown here) show normal behavior i.e. mode frequency decreases with increasing temperature. (ii) Linewidth of modes S6 ( $A_{1g}$ ), S7, S11, S20 and S1/S12/16 (not shown here) show normal temperature dependent i.e. linewidth increases with increasing temperature. Linewidth of S2 ($E_{2g}^1$) and S10 show anomalous temperature dependent behavior i.e. linewidth decreases with increasing temperature. Linewidth of modes S15 (not shown here) and S17 show non-monotonic behavior as a function of temperature. Linewidth of mode S19 shows normal behavior in the temperature range of 4 to ~ 240 K and then it shows a drop around this temperature and with further increase in temperature it shows increasing behavior till the highest recorded temperature (330 K).

First we will focus on understanding the frequency shift of the phonon modes as a function of temperature for few layer as well as 3L $MoS_2$. In the literature [50-53], a linear approximation, $\omega(T) = \omega_0 + \chi T$, has been employed to understand the temperature dependence of the mode frequencies, where $\omega_0$ is the frequency at 0 K, and $\chi$ is the first order temperature coefficient. However, we note that this linear approximation works well only at high temperature and does not hold good at low temperature especially below ~ 150 K. From the Figs. 4 (a) and 4 (c) for few layers, and Figs. 5 (a) and 5 (c) for 3L $MoS_2$, one can see that frequencies of the modes become nearly constant below a certain temperature which lies in between 80 to 160 K, and above 160 K the frequencies of the phonon modes decrease with increasing temperature. Therefore, linear approximation can not be used to understand the temperature dependence of phonon frequency shifts in the whole experimental temperature range (i.e. 4 to 330 K). In order to understand the temperature dependence of phonon frequencies in the whole experimental temperature range, a physical model including three and four phonon anharmonic effect may be used to analyze the temperature dependence of the phonon modes frequencies. The temperature dependence of the phonon frequency attributed to anharmonic effect may be expressed as $\omega(T) = \omega_0 + \Delta\omega_{anh}$, where $\omega_0$ is the frequency at 0 K, and $\Delta\omega_{anh}$ is the change in frequency due to anharmonic effect at finite temperature. Temperature dependence of the phonon frequency due to anharmonic effect arises from the phonon-phonon interaction via three and four phonon process [54-55] and may be expressed as

$$\omega(T) = \omega(0) + A[1 + \frac{2}{e^x - 1})] + B[1 + \frac{3}{e^y - 1} + \frac{3}{(e^y - 1)^2}] \quad \text{----------} (1)$$

Where $x = (\hbar\omega_0 / 2kT)$ and $y = (\hbar\omega_0 / 3kT)$ correspond to three and four phonon decay process contributing to the change in the phonon frequency with temperature. In three phonon anharmonic process, an optical phonon decay into two phonons with equal phonon frequency i.e. $(\omega_1 = \omega_2)$ and opposite momentum $(k_1 + k_2 = 0)$; whereas in four phonon anharmonic process, an optical phonon decays into three phonons $(\omega_1 = \omega_2 = \omega_3; k_1 + k_2 + k_3 = 0)$. The coefficient A and B are the self-energy fitting parameter, representing the contributions due to three and four phonon process to the frequency shift, respectively. In Figs. 4 (a) and 4 (c) and Figs. 5 (a) and 5 (c), the solid red lines are the fitted curves using above eq$^n$ 1; extracted values of the fitted parameters are listed in Table-II and Table-III for few layers and 3L $MoS_2$, respectively. For the higher temperature, taking only the first term of the Taylor expansion (i.e. $e^t = 1 + t$), eq$^n$ 1 may be converted to a linear temperature dependence of the phonon modes. We have also extracted the linear temperature coefficient ($\chi$) for $E^1_{2g}$ and $A_{1g}$ modes by fitting of phonon frequency in a temperature window of 160-330 K. The extracted first order coefficient values of the mode S2 ($E^1_{2g}$) and S6 ($A_{1g}$) are -0.0103± 0.0003 ($cm^{-1} K^{-1}$), -0.008 ± 0.0004 ($cm^{-1} K^{-1}$); -0.0128 ± 0.0005 ($cm^{-1} K^{-1}$), -0.0150 ±0.002 ($cm^{-1} K^{-1}$) for few layers and 3L $MoS_2$, respectively. We note that the estimated first order temperature coefficient is very close to the values reported in earlier reports on few layers $MoS_2$ [50-51].

Now we focus on to understand the temperature dependence of the linewidth of the phonon modes. Linewidth of the phonon modes are also quantitatively understood by three and four phonon anharmonic model. Generally, the linewidth of the phonon mode decreases with decreasing temperature and the linewidth is inversely proportional to life time of phonon. At low temperature, low population of the phonons results in reduced phonon-phonon interaction which leads to increase in life-time of phonon and as a result decrease in linewidth. As the temperature increases, population of the phonons also increase resulting in a large phonon-phonon interaction, which is reflected as reduced life-time of the phonons. As a consequence, the linewidth of the phonons increase with increasing temperature. Temperature dependence linewidth of the phonon modes except few modes in spectral range of 360-700 $cm^{-1}$, considering three and four phonon anharmonic approximation may be given by expression as [54-55]

$$\Gamma(T) = \Gamma(0) + C\,[1 + \frac{2}{e^{x} - 1})] + D\,[1 + \frac{3}{e^{y} - 1} + \frac{3}{(e^{y} - 1)^{2}}] \qquad \text{---------- (2)}$$

Where coefficient C and D are the self-energy fitting parameter, correspond to strength of phonon-phonon interaction involving the three and four phonon process, respectively. In Figs. 4 (b) and 4 (d) and Figs. 5 (b) and 5 (d), the solid red lines are fitted the curves using above $eq^{n}$ 2, and the solid blue lines are guide to the eye. The extracted values of the fitted parameters are listed in Table-II and Table-III for few layers and 3L $MoS_2$, respectively. Fitting using the above equation is in very good agreement with the experimental data, suggesting the finite role of three and four phonon process.

### 3.4. Temperature dependent Raman scattering intensity of the phonon modes

Figures 6 (a) and 6 (b) show the Raman spectra in 3D wall view in the temperature range of 4 to 330 K for few layers (FL) and 3L $MoS_2$, respectively. For both FL and 3L $MoS_2$, the intensity of the Raman spectrum shows a clear temperature variation, also reflected in the color contour map (see Figs. 3 (c) and 3 (d)). Figure 6 (c) shows the temperature dependence of the normalized intensity of few prominent phonon modes S2, S6, S7, and S10 in the FL $MoS_2$. Following observations can be made: (i) The mode S2 ($E_{2g}^1$) shows mild increase in the intensity with decreasing temperature down to ~ 160 K, starting from 330 K, and on further decreasing the temperature an anomalous sharp increase in the intensity is observed. (ii) The intensity of S6 ($A_{1g}$) phonon mode remain nearly constant with decreasing temperature down to ~ 160 K, and below ~ 160 it shows sharp increase in intensity similar to S2 (iii) Interestingly, the intensity of S7 and S10 ($2LA(M)$) modes shows anomalous decrease with decreasing temperature down to ~ 180 K, below ~ 180 K nearly temperature independent behavior is observed till 140 K, and on further decreasing temperature the intensity increases drastically. Figure 6 (d) shows the temperature dependence of the normalized intensity of the S2, S6, S7 and S10 phonon modes for the case of 3L $MoS_2$. Following observations can be made: (i) The intensity of the mode S2 ($E_{2g}^1$) varies slowly with decreasing temperature down to ~ 200 -160 K starting from 330 K. On the other hand, intensity of the mode S6 ($A_{1g}$), first decrease slightly from 300 K to ~ 250 K ( see inset in Fig. 6 (d) ), and then it increases slowly till ~160 K ; and on further cooling below ~ 160 K , the intensity for the both modes (i.e. S2 and S6 ) increase drastically and reach a maximum value at ~ 40 K, and interestingly below 40 K, the intensity show slight decrease down to the lowest recorded temperature, 4 K ( see green shaded region in Fig. 6 (d)). Also above 300 K, the decrease in intensity is observed for the case of S6 mode (see inset in Fig. 6 (d)). (ii) The intensity of the mode S7 and S10 ($2LA(M)$) decrease with decrease in temperature form 330 to ~250 K ( see insets in Fig. 6 (d) ), then it remains nearly constant till ~160 K; upon further decrease in temperature below 160 K, a sharp increase in intensity is observed.

In order to understand the temperature dependence of the intensity of the phonon mode, quantum mechanical expression for the Raman scattering intensity may be used. In quantum mechanical picture, the Raman scattering is a three steps process; (1) absorption of a photon (2) creation of a phonon, and (3) emission of a photon. The Raman intensity of phonon mode considering the quantum mechanical picture may be given as [56]

$$I \propto \left| \frac{\langle g | H_{el-p} | i' \rangle \langle i' | H_{el-ph} | i \rangle \langle i | H_{el-p} | g \rangle}{(\Delta E_{ig} - E_L)(\Delta E_{i'g} - E_S)} \right|^2 \qquad \text{---------- (3)}$$

where $|i\rangle$ , $|i'\rangle$ and $|g\rangle$ are the intermediate, and ground states. $H_{el-p}$ and $H_{el-ph}$ are the electron-photon and electron-phonon interaction Hamiltonians, respectively. Matrix elements $\langle i | H_{el-p} | g \rangle$, $\langle g | H_{el-p} | i' \rangle$ and $\langle i' | H_{el-ph} | i \rangle$ correspond to the electron-photon (absorption), the electron-photon (emission), and the electron-phonon interaction, respectively. $\Delta E_{ig} = (E_i - E_g + i\gamma_1)$ represents the energy difference between the intermediate state $E_i$ and the ground states $E_g$, and $\Delta E_{i'g} = (E_{i'} - E_g + i\gamma_2)$ corresponds to the energy separation between the intermediate states $E_{i'}$ and the final/ground states $E_g$. $\gamma_1$ and $\gamma_2$ are the linewidth or damping constants, and are related to the finite lifetime of the intermediate states. The energy $E_L = \hbar\omega_L$, and $E_S = \hbar\omega_s = \hbar\omega_L - \hbar\omega_{ph}$ are the energy of the incoming photon and outgoing photon, respectively; where $\omega_L$ ($\omega_s$) is the frequency of the incident (scattered) photon, and $\omega_{ph}$ is the frequency of the corresponding phonon mode. If the intermediate states are real excitonic energy states rather than virtual states, then $E_i - E_g$ and $E_{i'} - E_g$ could represent the excitonic energy band gap, and in the case when the energy of the incident (scattered) photon is very close to $E_i - E_g$ ($E_{i'} - E_g$), the Raman scattering intensity will enhanced significantly because of resonance effect. For the case, if 1s is the main intermediate excitonic state, and that contributes to the Raman scattering in TMDC kind of 2D materials, the Raman scattering intensity expression given by eq$^n$ 3 above, may be simplified and is given as [56]

$$I \propto \left| \frac{1}{(E(T) - E_L + i\gamma(T))(E(T) - E_s + i\gamma(T))} \right|^2 \qquad \text{---------- (4)}$$

where $E(T)$ and $\gamma(T)$ are temperature dependent transition energies and damping constants of exciton, respectively. Here, all the matrix element in eq$^n$ 3 are considered constant. In the vicinity of resonance, the temperature dependent Raman intensity of the phonon modes is significantly affected by the tuning of the resonance effect, while on the other hand, under non-resonance condition, the Bose-Einstein factor play an important role in the temperature dependence of Raman intensity of the phonon modes.

We observed a rapid increase in intensity for all the modes in both few layers as well as 3L $MoS_2$ with decreasing temperature below ~200 -160 K, see yellow shaded region in Figs. 6 (c) and 6 (d); while, the intensity remains either constant or weakly dependent on temperature or decrease with decrease in temperature from 300 to 200 K. Our temperature dependent variations in intensity as a function of temperature clearly evidenced the tuning of resonance effect with lowering the temperature. There are different ways to tune the resonance condition either by changing the temperature/pressure or by changing the incident photon energy. Here resonance condition may be tuned via varying the temperature. We observed two excitonic features in the PL spectra as mentioned in earlier section, corresponding to the A and B exciton, and is centered at ~ 1.82 eV ; 1.83 eV and 1.96 eV; 1.97 eV for the few layer and 3L $MoS_2$, respectively (see Fig. 1 (d) ). The laser used in the present study is 633 nm (1.96 eV) which is close to the energy of B exciton at room temperature, resulting in the resonance effect due to B exciton. It is reported in literature for TMDC systems [57-59] that energy of the excitons increases with decreasing temperature. As the temperature decrease, the energy of B exciton starts deviating from the incident photon energy in the present case, and this may explain the decrease in intensity with decreasing temperature from 300 to ~200 -160 K ; while on the other hand, the energy of the A exciton is approaching towards incident photon energy, therefore A exciton starts resonating with incident photon energy, and this may explain the rapidly enhancement in intensity of the phonon modes with decreasing temperature, especially below 160 K, reflecting a prominent role of A exciton in the low temperature range. Further, we note that the intensity of S2 and S6 modes for 3L $MoS_2$, reach a maximum value at 40 K with decreasing temperature from160 K, and below 40 K it starts showing decreasing trend till 4K, see Fig. 6 (d). The decrease in intensity at very low temperature, especially below 40 K, for the case of 3L $MoS_2$, may be due to the change in sample morphology via forming the wrinkles or ripples in system due to thermal expansion coefficient (TEC) mismatch between $MoS_2$ film and substrate. Such decrease in intensity below 40 K, is not observed for the case of few layers $MoS_2$, and it may be understood as TEC mismatch between $MoS_2$ film will be reduced with increasing the number of layers, and this may give continuous increase in the intensity even below 40 K for the case of few layers $MoS_2$.

## 4. Summary and Conclusions

In summary, we have performed a comprehensive temperature dependent Raman study for 3L and few layers $MoS_2$ in a broad spectral and wide temperature range under resonance condition. Our report is a first comprehensive temperature dependence studies covering phonon dynamics of first-order as well as overtone/combination mode under resonance condition. The temperature dependence of the phonon frequency and FWHM in the entire temperature range (4 K to 330 K) is understood by three and four phonon anharmonic process. Davydov splitting of in-plane $E_{2g}^{1}$ mode and out-of-plane $A_{1g}$ mode is observed under resonance excitation condition. We observed that number of Davydov splitting is more in few layers $MoS_2$ as compared 3L $MoS_2$ indicating Davydov splitting increases with increasing layers. We also observed the tuning of resonance effect with change in temperature. Our results provide crucial insight to understand the temperature evolution of the phonon modes under resonance condition in $MoS_2$, and we believe that this will motivate more studies on other TMDCs under resonance condition to understand their underlying intriguing properties.

**Acknowledgement:** PK acknowledge DST India for the financial support and IIT Mandi for the experimental facilities.

**Table- I:** List of the experimentally observed phonon modes using with 633 nm laser along with their symmetry in few layers and 3L $MoS_2$ at 4 K and 300 K.

| Modes | FL-$MoS_2$ $\omega(4K)$ ($cm^{-1}$) | FL-$MoS_2$ $\omega(300K)$ ($cm^{-1}$) | 3L-$MoS_2$ $\omega(4K)$ ($cm^{-1}$) | 3L-$MoS_2$ $\omega(300K)$ ($cm^{-1}$) | Symmetry |
|---|---|---|---|---|---|
| **S1** | 380.8±0.3 | 378.2±0.3 | 380.5±0.2 | 378.5±0.2 | $E^2_{1u}(\Gamma)$ |
| **S2** | 384.9±0.1 | 382.7±0.1 | 385.6±0.1 | 383.5±0.1 | $E^1_{2g}(\Gamma)$ |
| **S3** | 386.5±0.2 | 384.4±0.1 | - | - | - |
| **S4** | 405.0±0.2 | 403.6±0.4 | 405.2±0.2 | - | $A_{1g}(R_2)$ |
| **S5** | 407.2±0.2 | - | - | - | $A_{1g}(R_2)$ |
| **S6** | 410.0±0.1 | 407.7±0.1 | 408.8±0.1 | 406.2±0.2 | $A_{1g}(R_1)$ |
| **S7** | 421.5±0.8 | 416.2±0.1 | 424.5±0.1 | 417.6±0.1 | $B^2_{2g}+E^2_{1u}(\Gamma)$ |
| **S8** | 425.2±0.3 | - | - | - | - |
| **S9** | 442.8±1.5 | 441.3±0.9 | 443.5±1.2 | 442.6±0.8 | $E^2_{2g}(\Gamma)+A_{1g}(\Gamma)$ |
| **S10** | 456.8±0.3 | 452.6±0.2 | 456.5±0.4 | 453.0±0.2 | $2LA(M)$ |
| **S11** | 465.8±0.2 | 461.0±0.2 | 465.8±0.3 | 460.3±0.2 | - |
| **S12** | 470.5±0.2 | 466.8±0.1 | 470.5±0.1 | 466.4±0.2 | $A_{2u}(\Gamma)$ |
| **S13** | 528.7±0.3 | 528.3±0.5 | 527.9±0.1 | 527.3±0.3 | $E_{1g}(M)+LA(M)$ |
| **S14** | 545.4±0.4 | 541.8±0.6 | - | - | - |
| **S15** | 562.0±1.8 | 557.2±1.4 | 562.4±1.2 | 560.2±1.2 | - |
| **S16** | 573.3±0.1 | 567.8±0.5 | 573.3±0.3 | 568.9±0.3 | $2E_{1g}(\Gamma)$ |
| **S17** | 591.4±0.5 | 589.4±0.4 | 591.2±0.4 | 589.6±0.3 | - |
| **S18** | 602.9±0.1 | 597.9±0.2 | 602.6±0.1 | 599.9±0.3 | $E^1_{2g}(M)+LA(M)$ |
| **#** | - | - | 621.9±1.7 | - | - |
| **S19** | 635.8±0.4 | 629.5±0.2 | 637.3±0.3 | 629.8±0.2 | - |
| **S20** | 645.8±0.1 | 641.5±0.1 | 646.1±0.1 | 640.9±0.2 | $A_{1g}(M)+LA(M)$ |

**Table- II:** List of the fitting parameters for the phonon modes in few layers $MoS_2$, fitted by using equation as described in the text. The units are in $cm^{-1}$.

| **Modes** | ω(0) | **A** | **B** | Γ(0) | **C** | **D** |
|---|---|---|---|---|---|---|
| **S1** | 382.7±0.2 | -2.1±0.2 | 0.04±0.05 | 4.2±0.5 | 1.6±0.5 | 0.3±0.1 |
| **S2** | 386.3±0.1 | -1.5±0.1 | -0.02±0.03 | 1.7±0.02 | 0.9±0.08 | -0.2±0.003 |
| **S3** | 387.3±0.1 | -0.7±0.1 | -0.2±0.03 | 1.3±0.09 | 0.1±0.1 | -0.03±0.02 |
| **S4** | 405.8±0.1 | -0.9±0.0 | 0.04±0.05 | 2.1±0.2 | 0.4±0.1 | 0.3±0.2 |
| **S5** | 408.0±0.1 | -0.9±0.1 | 0.007±0.1 | 3.5±0.1 | -0.5±0.3 | 0.7±0.2 |
| **S6** | 412.5±0.2 | -2.8±0.2 | 0.2±0.06 | 1.6±0.7 | 1.0±0.7 | 0.06±0.1 |
| **S7** | 425.2±0.1 | -4.5±0.1 | 1.0±0.02 | 11.7±0.9 | -3.1±0.9 | 1.5±0.2 |
| **S8** | 426.5±0.2 | -0.3±0.3 | -1.2±0.3 | - | - | - |
| **S9** | 442.9±0.3 | 0.1±0.04 | -0.4±0.04 | 16.1±1.4 | 3.8±1.5 | -0.9±0.3 |
| **S10** | 457.3±0.6 | -0.3±0.1 | -0.8±0.06 | 14.2±0.6 | -0.9±0.7 | 0.07±0.2 |
| **S11** | 466.9±0.5 | -1.5±0.2 | -0.6±0.1 | - | - | - |
| **S12** | 471.2±1.0 | -0.2±1.3 | -0.8±0.3 | 1.3±0.1 | 3.8±0.09 | -0.6±0.07 |
| **S13** | - | - | - | - | - | - |
| **S14** | 547.8±0.2 | -2.4±0.8 | -0.4±0.1 | - | - | - |
| **S15** | 570.5±0.5 | -7.4±0.8 | -0.04±0.4 | - | - | - |
| **S16** | 577.4±1.2 | -2.3±1.8 | -1.7±0.5 | 8.8±2.9 | -4.2±4.1 | 2.8±1.3 |
| **S17** | 592.0±1.4 | -0.3±1.9 | -0.7±0.4 | - | - | - |
| **S18** | 608.6±0.8 | -4.6±1.1 | -0.9±0.3 | - | - | - |
| **S19** | 638.7±0.5 | 0.8±0.4 | -3.9±0.5 | - | - | - |
| **S20** | 648.0±1.5 | 0.7±2.1 | -2.8±0.7 | - | - | - |

**Table- III:** List of the fitting parameters for the phonon modes in 3L MoS2, fitted by using equation as described in the text. The units are in $cm^{-1}$.

| Modes | ω(0) | A | B | Γ(0) | C | D |
|---|---|---|---|---|---|---|
| **S1** | 381.5±0.2 | -0.3±0.2 | -0.3±0.05 | 8.0±0.5 | -2.0±0.5 | 0.7±0.1 |
| **S2** | 385.8±0.1 | 0.3±0.1 | -0.4±0.03 | 2.5±0.1 | 0.3±0.1 | -0.1±0.04 |
| **S4** | 406.3±0.1 | -1.1±0.2 | 0.05±0.07 | 2.9±0.4 | -0.2±0.06 | -0.04±0.03 |
| **S6** | 409.9±0.7 | -0.8±0.9 | -0.3±0.2 | 1.8±0.6 | 0.6±0.8 | 1.7±0.2 |
| **S7** | 425.9±0.4 | -1.2±0.2 | -1.0±0.1 | 6.4±0.6 | 0.1±0.8 | 0.8±0.2 |
| **S9** | 443.4±1.7 | 0.2±2.1 | -0.3±0.5 | - | - | - |
| **S10** | 456.7±0.6 | 0.7±0.7 | -1.1±0.1 | 13.8±0.9 | -1.5±0.1 | -0.2±0.07 |
| **S11** | 469.7±0.9 | -3.0±1.0 | -0.8±0.2 | 5.2±0.7 | 2.2±0.8 | -0.2±0.2 |
| **S12** | 473.8±0.4 | -2.8±0.5 | -0.5±0.1 | 2.6±.01 | 1.2±0.11 | 0.3±0.06 |
| **S13** | - | - | - | - | - | - |
| **S15** | 566.9±0.9 | -4.3±0.6 | 0.02±0.4 | - | - | - |
| **S16** | 575.9±0.9 | -1.0±1.3 | -1.6±0.4 | 4.2±0.7 | 0.6±0.3 | 0.8±0.2 |
| **S17** | 595.9±1.1 | -5.1±2.5 | 0.7±0.7 | - | - | - |
| **S18** | 604.1±1.1 | -0.4±1.1 | -1.3±0.5 | 7.2±0.8 | 1.9±0.6 | -1.6±0.5 |
| **S19** | 639.3±0.5 | 0.4±0.3 | -3.6±0.2 | - | - | - |
| **S20** | 649.1±0.5 | -1.1±0.1 | -2.2±0.1 | 3.4±2.0 | 3.5±2.8 | -0.02±0.9 |

## FIGURE CAPTION:

**FIGURE 1:** (Color online) (a) optical micrograph for the horizontally aligned CVD grown layered $MoS_2$. The region where Raman spectra were measured are labeled by numbers (1) and (2) corresponds to flake 1 and flake 2. (b) Raman spectra for flake 1 and flake 2 excited with 532 nm at room temperature. (c) Schematic representation of the excitonic energy level. (d) PL spectra collected from the flake 1 and flake 2 excited with 532 nm at room temperature. $X_A$ and $X_B$ marked the position of A and B exciton energy.

**FIGURE 2:** (Color online) (a) and (b) Raman spectra of the horizontally aligned CVD grown 3L and few layers (FL) $MoS_2$, respectively, recorded at 300 K using 633 nm laser in the spectral range of 360-700 $cm^{-1}$. The solid red line shows the total sum of Lorentizian fit and thin blue lines show the individual fit of the phonon modes. Solid black sphere are experimental data points.

**FIGURE 3:** (Color online) Temperature evaluation of the Raman spectra showing Davydov splitting of the modes for (a) few layers (FL) and (b) 3L $MoS_2$ in the spectral range of 360-435 $cm^{-1}$. Solid red line shows the total sum of Lorentzian fit and thin blue lines show individual fit of the phonon modes. Solid black sphere are experimental data point. Dark-shaded region depicts the emergence of new modes as a result of Davydov splitting at low temperature. (c) and (d) 2D color contour maps of the Raman intensity versus Raman shift and as a function of temperature for few layers (FL) and 3L $MoS_2$, respectively.

**FIGURE 4:** (Color online) (a) and (c) Temperature dependence of frequency of the phonon modes S2, S4-S6, S7, S8, S12, S16, S18, S19 and S20; (b) and (d) Temperature dependence of the full width at half maximum (FWHM) of the phonon modes S2, S6, S7, S12, S16, S18, S19 and S20 for few layers (FL) $MoS_2$. Solid red lines are the fitted curve as described in the text and solid blue lines are guide to the eye.

**FIGURE 5:** (Color online) (a) and (c) Temperature dependence of frequency of the phonon modes S2, S4, S6, S7, S10, S11, S17, S19, and S20 ; (b) and (d) Temperature dependence of the full width at half maximum (FWHM) of the phonon modes S2, S6, S7, S10, S11, S17, S19 and S20 for 3L $MoS_2$. Solid red lines are the fitted curve as described in the text and solid blue lines are guide to the eye.

**FIGURE 6:** (Color online) (a) and (b) show the Raman spectrum in 3D wall view in the temperature range of 4 K to 330 K, in a spectral ranges from 330-700 $cm^{-1}$ for few layers (FL) and 3L MoS2, respectively. (c) and (d) show the temperature dependence of normalized intensity of phonon modes S2 ($E_{2g}^1$), S6 ($A_{1g}$), S7 and S10 ($2LA(M)$) for FL and 3L $MoS_2$, respectively. The yellow shaded part depicts the region where resonance effect is tuned from B exciton to A exciton. The inset figures in (d) show the temperature dependence of intensity in temperature range of 180 K to 330 K. Solid green lines are guide to the eye.

## FIGURE 1:

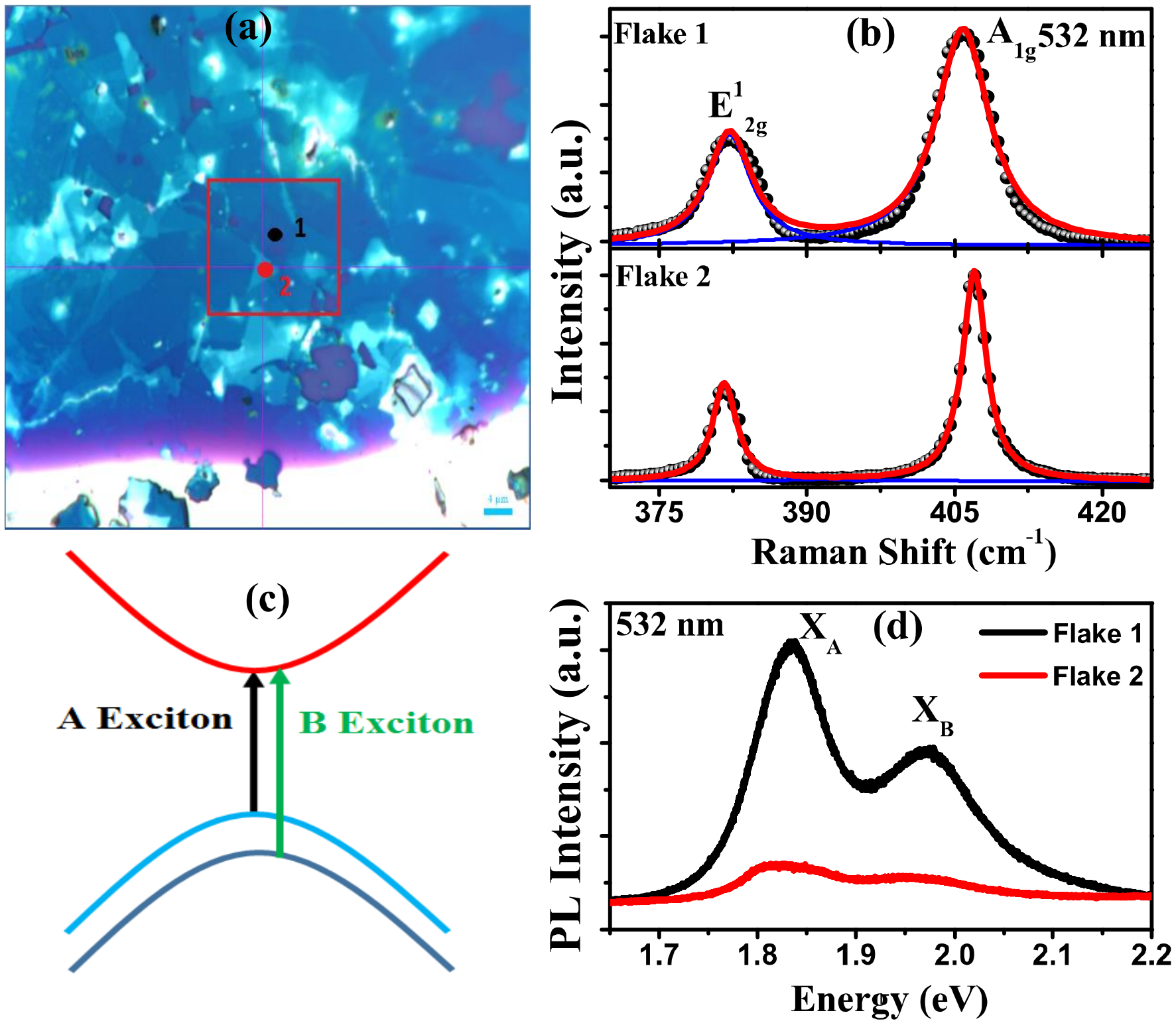

# FIGURE 2:

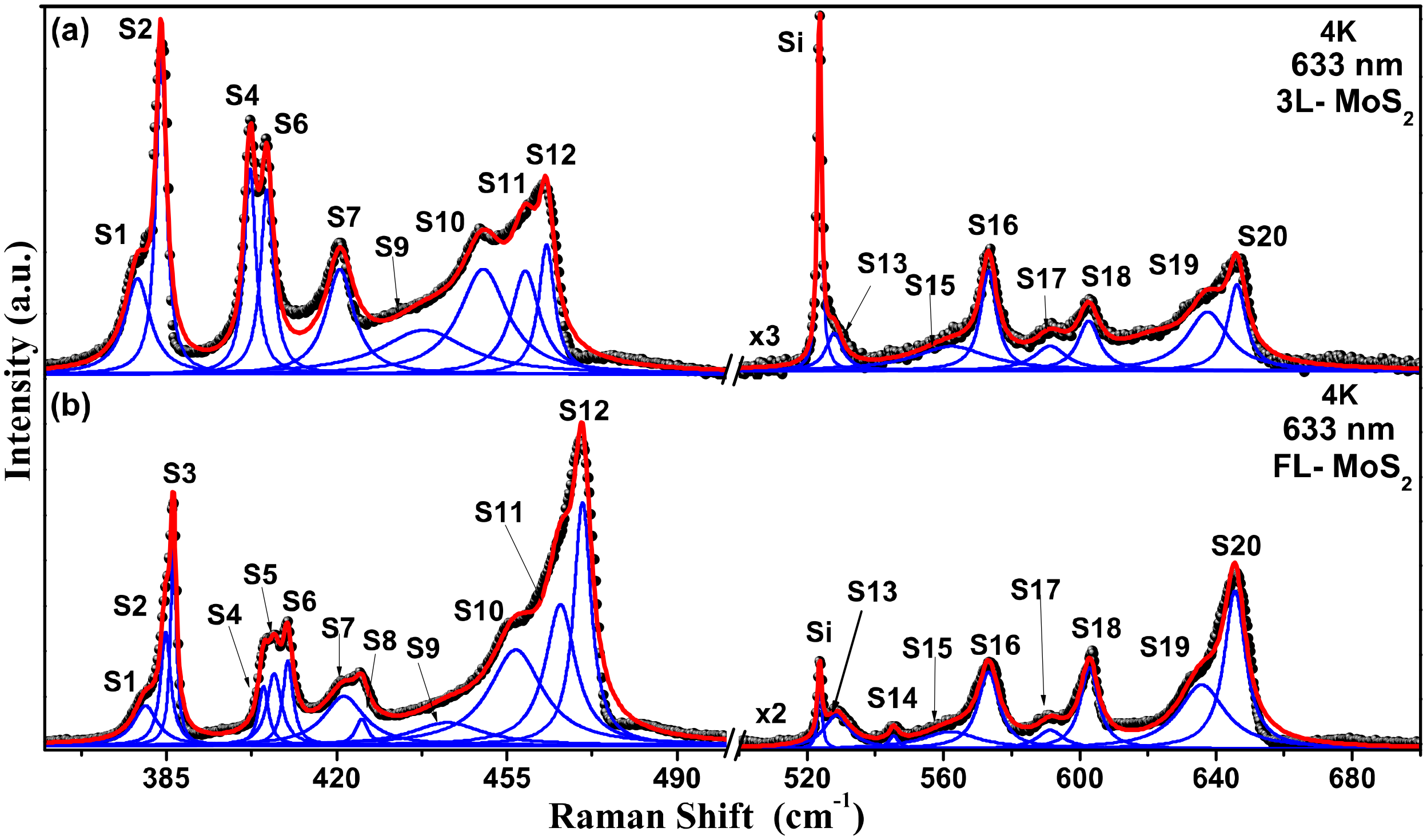

# FIGURE 3:

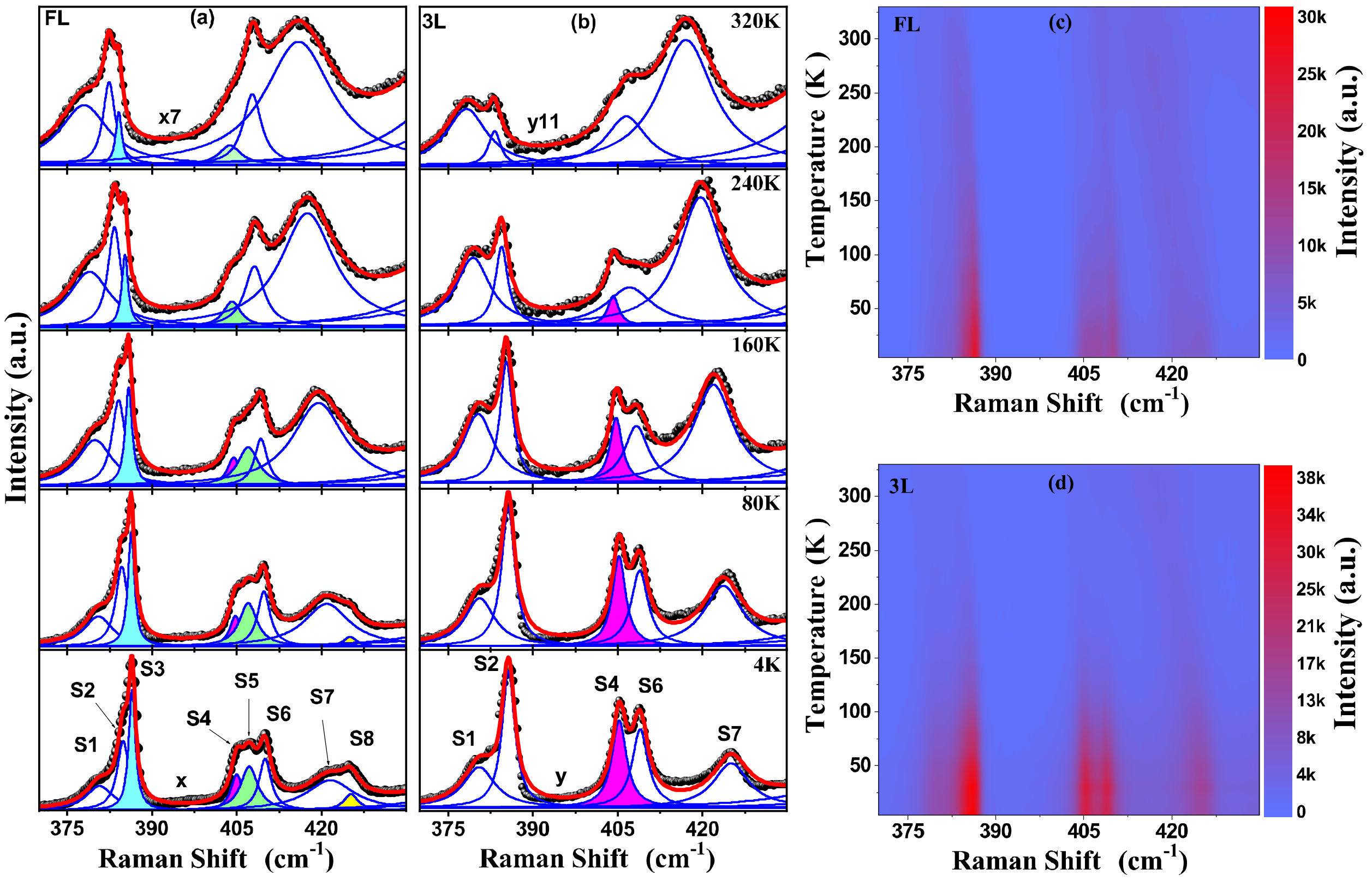

**FIGURE 4:**

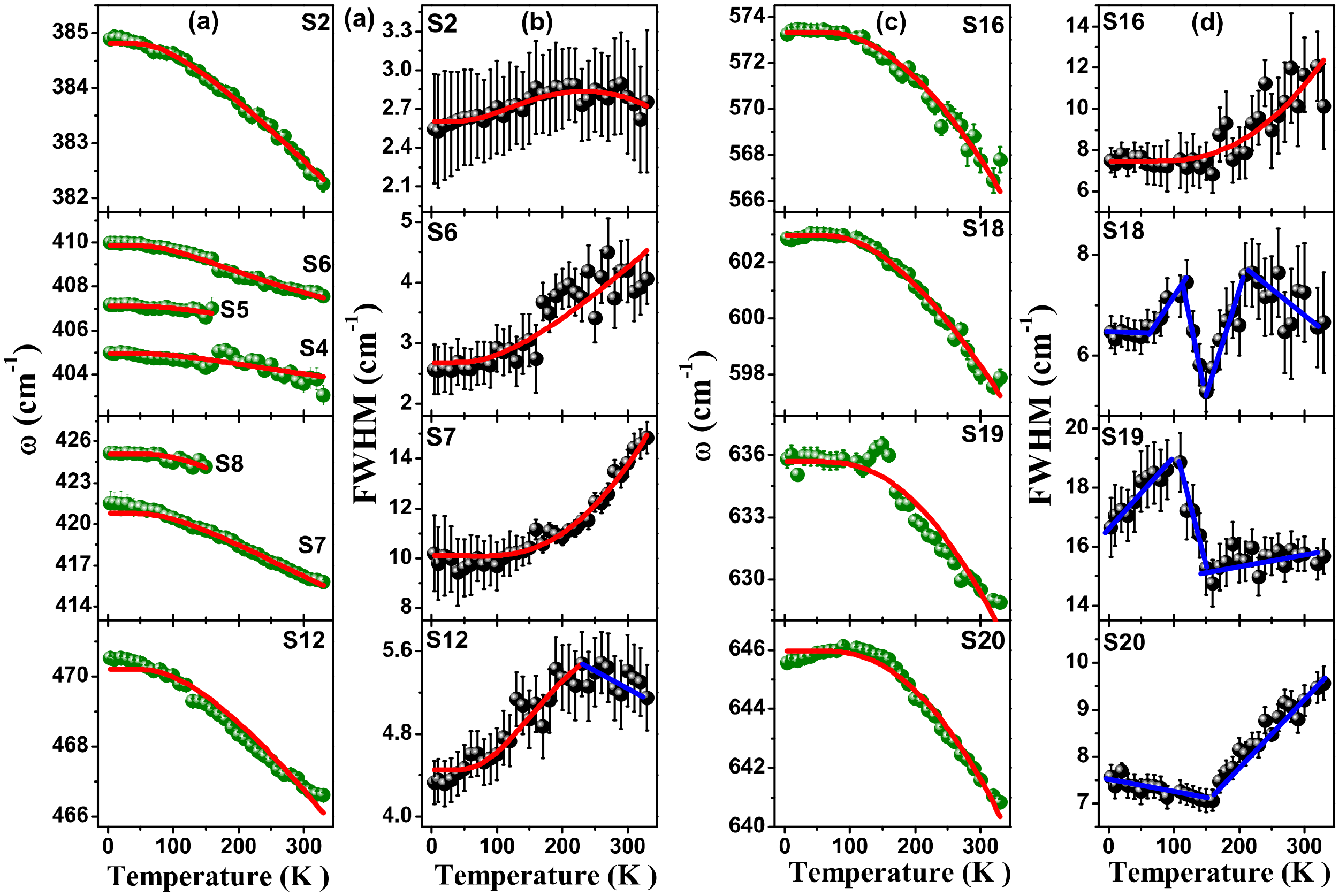

**FIGURE 5:**

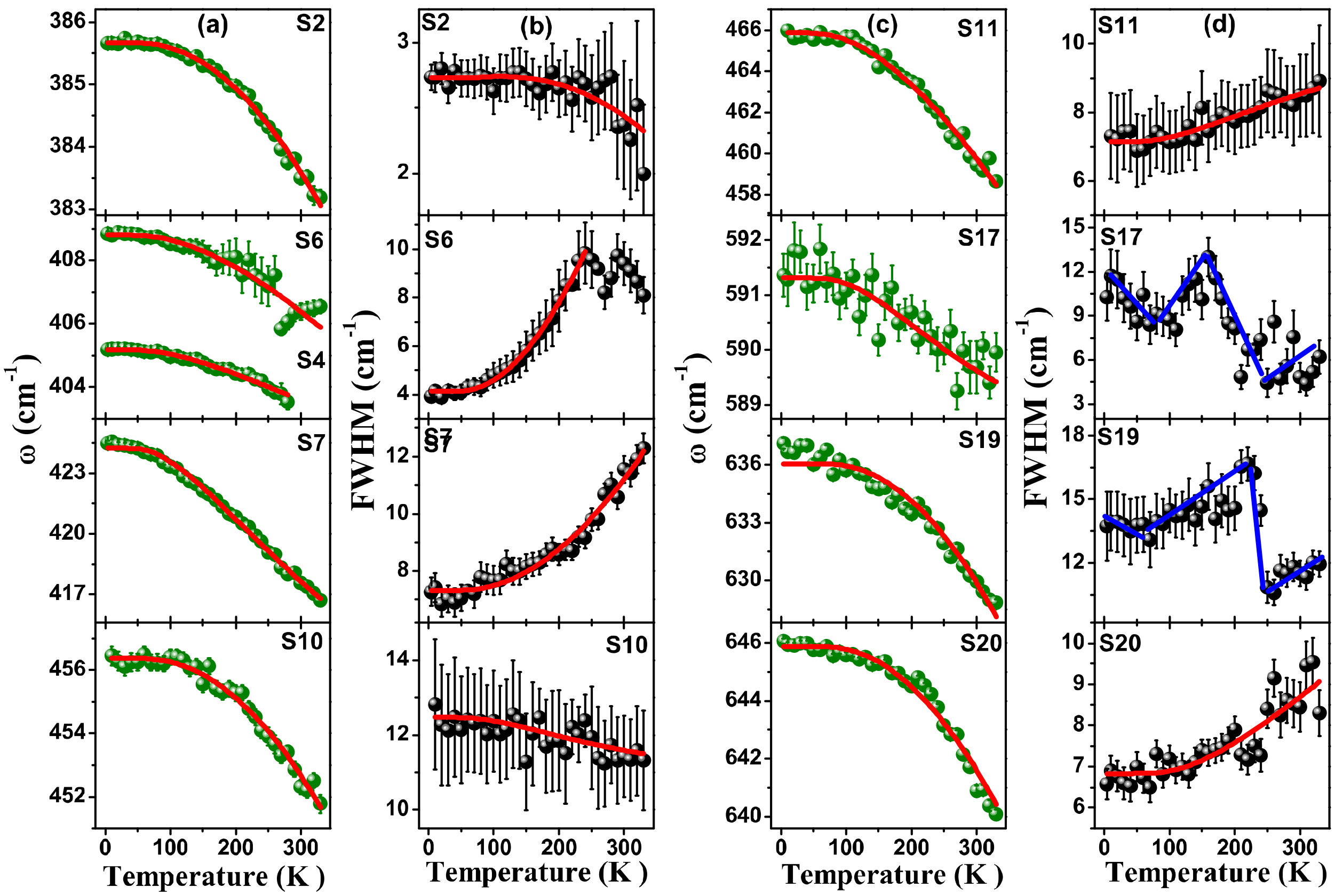

## FIGURE 6:

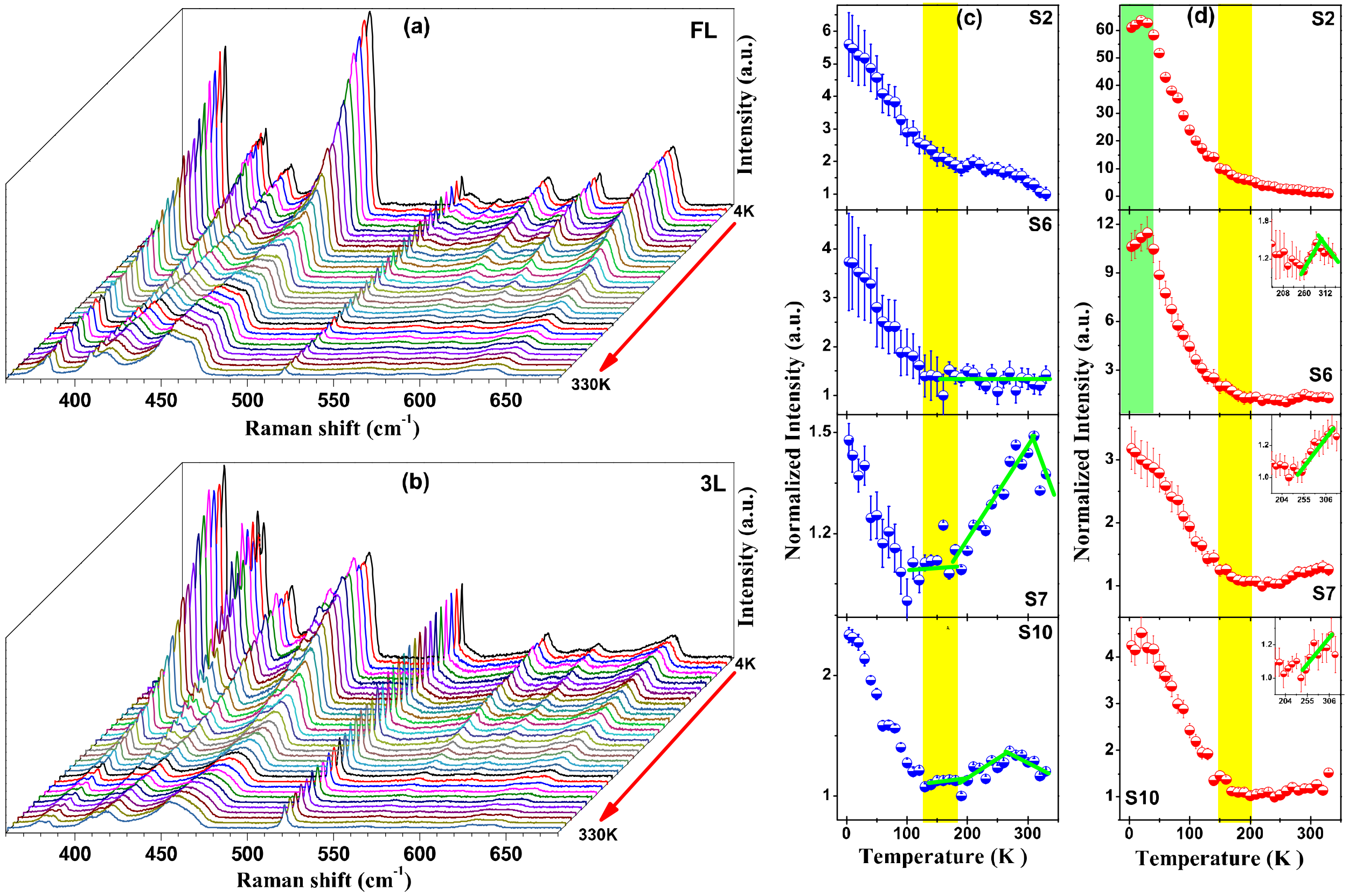